\documentclass[conference]{IEEEtran}
\ifCLASSINFOpdf
\else
   \usepackage[dvips]{graphicx}
   \newcommand{\bysame}{%
    \leavevmode\hbox to 3em{\hrulefill}\,}
\fi
\begin{document}
\title{Error-Trellis Construction for Tailbiting Convolutional Codes}

\author{\IEEEauthorblockN{Masato Tajima}
\IEEEauthorblockA{Graduate School of Science and Engineering\\
University of Toyama\\
3190 Gofuku, Toyama 930-8555, Japan\\
Email: tajima@eng.u-toyama.ac.jp}
\and
\IEEEauthorblockN{Koji Okino}
\IEEEauthorblockA{Information Technology Center\\
University of Toyama\\
3190 Gofuku, Toyama 930-8555, Japan\\
Email: okino@itc.u-toyama.ac.jp}
%\and
%\IEEEauthorblockN{Takashi Miyagoshi}
%\IEEEauthorblockA{Graduate School of Sci. and Eng.\\
%University of Toyama\\
%3190 Gofuku, Toyama 930-8555, Japan\\
%Email: miyagosi@eng.u-toyama.ac.jp}
}

%Telephone: (800) 555--1212, Fax: (888) 555--1212}


\maketitle

\begin{abstract}
%\boldmath
In this paper, we present an error-trellis construction for tailbiting convolutional codes. A tailbiting error-trellis is characterized by the condition that the syndrome former starts and ends in the same state. We clarify the correspondence between code subtrellises in the tailbiting code-trellis and error subtrellises in the tailbiting error-trellis. Also, we present a construction of tailbiting backward error-trellises. Moreover, we obtain the scalar parity-check matrix for a tailbiting convolutional code. The proposed construction is based on the adjoint-obvious realization of a syndrome former and its behavior is fully used in the discussion.
\end{abstract}


% \fi
\IEEEpeerreviewmaketitle

\section{Introduction}
In this paper, we always assume that the underlying field is $F=\mbox{GF}(2)$. Let $G(D)$ be a generator matrix of an $(n, k)$ convolutional code $C$. Let $H(D)$ be a corresponding $r \times n$ parity-check matrix of $C$, where $r=n-k$. Both $G(D)$ and $H(D)$ are assumed to be canonical [1], [5]. Denote by $L$ the memory length of $G(D)$ (i.e., the maximum degree among the polynomials of $G(D)$) and by $M$ the memory length of $H(D)$. Then $H(D)$ is expressed as
\begin{equation}
H(D)=H_0+H_1D+ \cdots +H_MD^M .
\end{equation}
Consider a terminated version of $C$ with $N$ trellis sections. That is, each codeword is a path starting from the all-zero state at time $t=0$ and ending in the all-zero state at time $t=N$. In this case, $C$ is specified by the following scalar parity-check matrix [1], [6]:
\begin{equation}
H_{scalar}=\left(
\begin{array}{ccccc}
H_0 &  &  &  &  \\
H_1 & H_0 &  &  &  \\
\scriptstyle{\ldots} & H_1 & \scriptstyle{\ldots} &  &  \\
\scriptstyle{\ldots} & \scriptstyle{\ldots} & \scriptstyle{\ldots} & \scriptstyle{\ldots} &  \\
H_M & \scriptstyle{\ldots} & \scriptstyle{\ldots} & \scriptstyle{\ldots} & H_0 \\
 & H_M & \scriptstyle{\ldots} & \scriptstyle{\ldots} & H_1 \\
 &  & \scriptstyle{\ldots} & \scriptstyle{\ldots} & \scriptstyle{\ldots} \\
 &  &  & \scriptstyle{\ldots} & \scriptstyle{\ldots} \\
 &  &  &  & H_M
\end{array}
\right)
\end{equation}
with size $(N+M)r\times Nn$ (blanks indicate zeros).
\begin{figure}[tb]
\begin{center}
\includegraphics[width=8.0cm,clip]{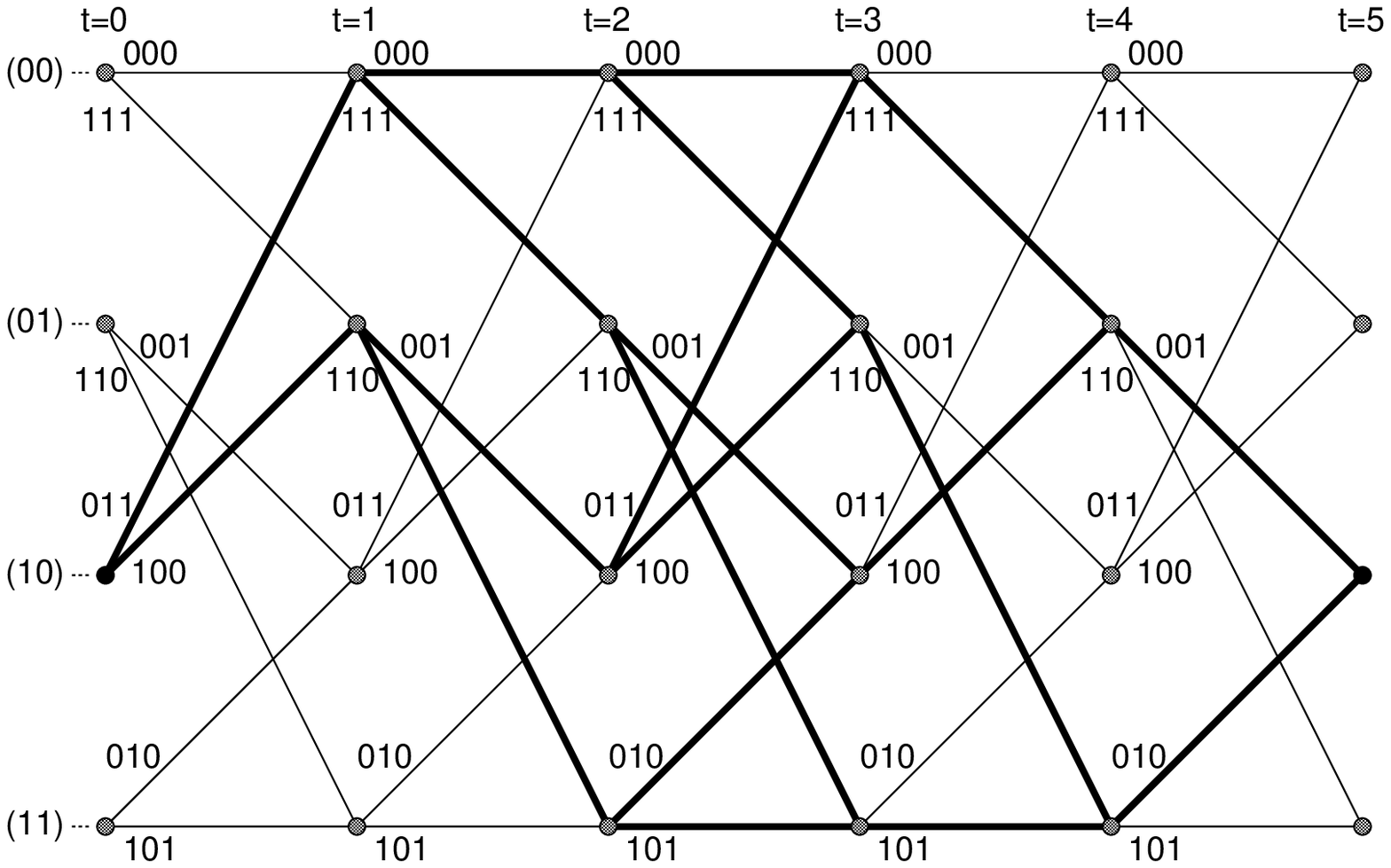}
\end{center}
\caption{Tailbiting code-trellis based on $G_1(D)$.}
\label{Fig.1}
\end{figure}
\par
Tailbiting is a technique by which a convolutional code can be used to construct a block code without any loss of rate [4], [7], [10]. Let $C_{tb}$ be a tailbiting convolutional code with an $N$-section code-trellis $T_{tb}^{(c)}$. The fundamental idea behind tailbiting is that the encoder starts and ends in the same state, i.e., $\mbox{\boldmath $\beta$}_0=\mbox{\boldmath $\beta$}_N$ ($\mbox{\boldmath $\beta$}_k$ is the encoder state at time $k$). Suppose that $T_{tb}^{(c)}$ has $\Sigma_0$ initial (or final) states, then it is composed of $\Sigma_0$ subtrellises, each having the same initial and final states. We call these subtrellises tailbiting code subtrellises. For example, a tailbiting code-trellis of length $N=5$ based on the generator matrix
\begin{equation}
G_1(D)=(1, 1+D^2, 1+D+D^2)
\end{equation}
is shown in Fig.1. Since $\Sigma_0=4$, this tailbiting code-trellis is composed of $4$ code subtrellises. In Fig.1, bold lines correspond to the code subtrellis with $\mbox{\boldmath $\beta$}_0=\mbox{\boldmath $\beta$}_5=(1,0)$.
\par
On the other hand, it is reasonable to think that an error-trellis $T_{tb}^{(e)}$ for the tailbiting convolutional code $C_{tb}$ can equally be constructed. In this case, each error subtrellis should have the same initial and final states like a code subtrellis. In this paper, taking this property into consideration, we present an error-trellis construction for tailbiting convolutional codes. We also clarify the correspondence between code subtrellises in $T_{tb}^{(c)}$ and error subtrellises in $T_{tb}^{(e)}$. In this relationship, we see that dual states (i.e., syndrome-former states corresponding to encoder states) play an important role. Also, a kind of superposition rule associated with a syndrome former is used. Next, we present a construction of tailbiting backward error-trellises. Using the backward error-trellis, each tailbiting error path is represented in time-reversed order. Moreover, we derive the general structure of the scalar parity-check matrix for a tailbiting convolutional code. Similar to a scalar generator matrix, it is shown that the obtained scalar parity-check matrix has a cyclic structure. In general, unlike code-trellises, error-trellises enable decoding with remarkably low average complexity [1]. Hence, we think an error-trellis construction presented in this paper is very important.

\section{Syndrome Former $H^T(D)$}
\subsection{Adjoint-Obvious Realization of a Syndrome Former}
Consider the adjoint-obvious realization (observer canonical form [2], [3]) of the syndrome former $H^T(D)$ ($T$ means transpose). Let $\mbox{\boldmath $e$}_k=(e_k^{(1)}, e_k^{(2)},\cdots, e_k^{(n)})$ and $\mbox{\boldmath $\zeta$}_k=(\zeta_k^{(1)}, \zeta_k^{(2)}, \cdots, \zeta_k^{(r)})$ be the input error at time $k$ and the corresponding output syndrome at time $k$, respectively. Denote by $\sigma_{kp}^{(q)}$ the contents of the memory elements in the above realization. Here, the contents of the memory array corresponding to the syndrome bit $\zeta_k^{(q)}$ are labeled with $q$. For any fixed $q$, $\sigma_{k1}^{(q)}$ corresponds to the memory element which is closest to the $q$th output of the syndrome former (i.e., $\zeta_k^{(q)}$). If a memory element is missing, the corresponding $\sigma_{kp}^{(q)}$ is set to zero. Using $\sigma_{kp}^{(q)}$, the syndrome-former state at time $k$ is defined as
\begin{equation}
\mbox{\boldmath $\sigma$}_k\stackrel{\triangle}{=}(\sigma_{k1}^{(1)}, \cdots, \sigma_{k1}^{(r)}, \cdots, \sigma_{kM}^{(1)}, \cdots, \sigma_{kM}^{(r)}) .
\end{equation}
({\it Remark:} The effective size of $\mbox{\boldmath $\sigma$}_k$ is equal to the overall constraint length of $H(D)$.)
\par
Let $\mbox{\boldmath $\xi$}_k\stackrel{\triangle}{=}(\mbox{\boldmath $\zeta$}_k, \mbox{\boldmath $\sigma$}_k)^T$ be the extended state augmented with the syndrome $\mbox{\boldmath $\zeta$}_k$. Then $\mbox{\boldmath $\xi$}_k$ has an expression [8], [9]:
\begin{eqnarray}
\mbox{\boldmath $\xi$}_k &=& \left(
\begin{array}{ccccc}
H_M & H_{M-1} & \scriptstyle{\ldots} & H_1 & H_0 \\
0 & H_M & \scriptstyle{\ldots} & H_2 & H_1\\
\scriptstyle{\ldots} & \scriptstyle{\ldots} & \scriptstyle{\ldots} & \scriptstyle{\ldots} & \scriptstyle{\ldots} \\
0 & 0 & \scriptstyle{\ldots} & H_M & H_{M-1} \\
0 & 0 & \scriptstyle{\ldots} & 0 & H_M
\end{array}
\right) \nonumber \\
&& \times \left(\mbox{\boldmath $e$}_{k-M}, \mbox{\boldmath $e$}_{k-M+1}, \cdots, \mbox{\boldmath $e$}_k\right)^T \nonumber \\
&\stackrel{\triangle}{=}& H^*\times\left(\mbox{\boldmath $e$}_{k-M}, \mbox{\boldmath $e$}_{k-M+1}, \cdots, \mbox{\boldmath $e$}_k\right)^T .
\end{eqnarray}
From this expression, we have
\begin{eqnarray}
\mbox{\boldmath $\sigma$}_k &\stackrel{\triangle}{=}& (\mbox{\boldmath $\sigma$}_k^{(1)}, \mbox{\boldmath $\sigma$}_k^{(2)}, \cdots, \mbox{\boldmath $\sigma$}_k^{(M)}) \nonumber \\
&=& (\mbox{\boldmath $e$}_{k-M+1}, \cdots, \mbox{\boldmath $e$}_{k-1}, \mbox{\boldmath $e$}_k) \nonumber \\
&& \times \left(
\begin{array}{cccc}
H_M^T & \scriptstyle{\ldots} & 0 & 0 \\
\scriptstyle{\ldots} & \scriptstyle{\ldots} & \scriptstyle{\ldots} & \scriptstyle{\ldots} \\
H_2^T & \scriptstyle{\ldots} & H_M^T & 0 \\
H_1^T & \scriptstyle{\ldots} & H_{M-1}^T & H_M^T
\end{array}
\right) \nonumber \\
&\stackrel{\triangle}{=}& (\mbox{\boldmath $e$}_{k-M+1}, \cdots, \mbox{\boldmath $e$}_{k-1}, \mbox{\boldmath $e$}_k) \times H^{**T} .
\end{eqnarray}
Note that $\mbox{\boldmath $\sigma$}_k$ has an alternative expression:
\begin{equation}
\mbox{\boldmath $\sigma$}_k=(\mbox{\boldmath $\sigma$}_{k-1}^{(2)}, \cdots, \mbox{\boldmath $\sigma$}_{k-1}^{(M)}, \mbox{\boldmath $0$})+\mbox{\boldmath $e$}_k(H_1^T, H_2^T, \cdots, H_M^T) .
\end{equation}
Similarly, $\mbox{\boldmath $\zeta$}_k$ is expressed as
\begin{eqnarray}
\mbox{\boldmath $\zeta$}_k &=& \mbox{\boldmath $e$}_{k-M}H_M^T+\cdots+\mbox{\boldmath $e$}_{k-1}H_1^T+\mbox{\boldmath $e$}_kH_0^T \\
&=& \mbox{\boldmath $\sigma$}_{k-1}^{(1)}+\mbox{\boldmath $e$}_kH_0^T .
\end{eqnarray}

\subsection{Dual States}
The encoder states can be labeled by the syndrome-former states (i.e., dual states [2]). The dual state $\mbox{\boldmath $\beta$}_k^*$ corresponding to the encoder state $\mbox{\boldmath $\beta$}_k$ is obtained by replacing $\mbox{\boldmath $e$}_k$ in $\mbox{\boldmath $\sigma$}_k$ by $\mbox{\boldmath $y$}_k=\mbox{\boldmath $u$}_kG(D)$ ($\mbox{\boldmath $u$}_k$ is the information at time $k$). We have
\begin{eqnarray}
\mbox{\boldmath $\beta$}_k^* &=& (\mbox{\boldmath $y$}_{k-M+1}, \cdots, \mbox{\boldmath $y$}_{k-1}, \mbox{\boldmath $y$}_k) \nonumber \\
&& \times \left(
\begin{array}{cccc}
H_M^T & \scriptstyle{\ldots} & 0 & 0 \\
\scriptstyle{\ldots} & \scriptstyle{\ldots} & \scriptstyle{\ldots} & \scriptstyle{\ldots} \\
H_2^T & \scriptstyle{\ldots} & H_M^T & 0 \\
H_1^T & \scriptstyle{\ldots} & H_{M-1}^T & H_M^T
\end{array}
\right) .
\end{eqnarray}
\par
{\it Example 1:} Consider the parity-check matrix
\begin{equation}
H_1(D)=\left(
\begin{array}{ccc}
1+D& D & 1+D \\
D& 1 & 1 
\end{array}
\right)
\end{equation}
corresponding to $G_1(D)$. $H_1(D)$ is expressed as
\begin{eqnarray}
H_1(D) &=& \left(
\begin{array}{ccc}
1& 0 & 1 \\
0& 1 & 1 
\end{array}
\right)+\left(
\begin{array}{ccc}
1& 1 & 1 \\
1& 0 & 0 
\end{array}
\right)D \nonumber \\
&\stackrel{\triangle}{=}& H_0+H_1D .
\end{eqnarray}
Hence ($M=1$), the dual state corresponding to the encoder state $\mbox{\boldmath $\beta$}_k=(u_{k-1}, u_k)$ is obtained as follows.
\begin {eqnarray}
\mbox{\boldmath $\beta$}_k^* &=& \mbox{\boldmath $y$}_kH_1^T \nonumber \\
&=& (y_k^{(1)}, y_k^{(2)}, y_k^{(3)}) \left(
\begin{array}{cc}
1 & 1 \\
1 & 0 \\
1 & 0 
\end{array}
\right) \nonumber \\
&=& (y_k^{(1)}+y_k^{(2)}+y_k^{(3)}, y_k^{(1)}) \nonumber \\
&=& (u_{k-1}+u_k, u_k) .
\end{eqnarray}

\subsection{Behavior of a Syndrome Former}
\newtheorem{lemm}{Lemma}
\begin{lemm}Let $\mbox{\boldmath $\sigma$}_{k-1}$ be the syndrome-former state at time $k-1$. Here, assume that an error $\mbox{\boldmath $e$}_k$ is inputted to the syndrome former and it moves to the state $\mbox{\boldmath $\sigma$}_k$ at time $k$. Also, assume that the syndrome $\mbox{\boldmath $\zeta$}_k$ is outputted according to this transition. (This relation is denoted as
$$
\mbox{\boldmath $\sigma$}_{k-1}\mathop{\longrightarrow}_{\mbox{\boldmath $\zeta$}_k}^{\mbox{\boldmath $e$}_k}\mbox{\boldmath $\sigma$}_k .\mbox{)}
$$
Similarly, assume the relation
\begin{equation}
\mbox{\boldmath $\sigma$}_{k-1}'\mathop{\longrightarrow}_{\mbox{\boldmath $\zeta$}_k'}^{\mbox{\boldmath $e$}_k'}\mbox{\boldmath $\sigma$}_k' .
\end{equation}
Then we have
\begin{equation}
\mbox{\boldmath $\sigma$}_{k-1}+\mbox{\boldmath $\sigma$}_{k-1}'\mathop{\longrightarrow}_{\mbox{\boldmath $\zeta$}_k+\mbox{\boldmath $\zeta$}_k'}^{\mbox{\boldmath $e$}_k+\mbox{\boldmath $e$}_k'}\mbox{\boldmath $\sigma$}_k+\mbox{\boldmath $\sigma$}_k' .
\end{equation}
\end{lemm}
\begin{IEEEproof}From the assumption, the relations
\begin{equation}
\mbox{\boldmath $\sigma$}_k=(\mbox{\boldmath $\sigma$}_{k-1}^{(2)}, \cdots, \mbox{\boldmath $\sigma$}_{k-1}^{(M)}, \mbox{\boldmath $0$})+\mbox{\boldmath $e$}_k(H_1^T, H_2^T, \cdots, H_M^T)
\end{equation}
\begin{equation}
\mbox{\boldmath $\sigma$}_k'=(\mbox{\boldmath $\sigma$}_{k-1}'^{(2)}, \cdots, \mbox{\boldmath $\sigma$}_{k-1}'^{(M)}, \mbox{\boldmath $0$})+\mbox{\boldmath $e$}_k'(H_1^T, H_2^T, \cdots, H_M^T)
\end{equation}
hold. Hence, we have
\begin{eqnarray}
\mbox{\boldmath $\sigma$}_k+\mbox{\boldmath $\sigma$}_k' &=& (\mbox{\boldmath $\sigma$}_{k-1}^{(2)}+\mbox{\boldmath $\sigma$}_{k-1}'^{(2)}, \cdots, \mbox{\boldmath $\sigma$}_{k-1}^{(M)}+\mbox{\boldmath $\sigma$}_{k-1}'^{(M)}, \mbox{\boldmath $0$}) \nonumber \\
&& +(\mbox{\boldmath $e$}_k+\mbox{\boldmath $e$}_k')(H_1^T, H_2^T, \cdots, H_M^T) .
\end{eqnarray}
On the other hand, using the relations
\begin{eqnarray}
\mbox{\boldmath $\zeta$}_k &=& \mbox{\boldmath $\sigma$}_{k-1}^{(1)}+\mbox{\boldmath $e$}_kH_0^T \\
\mbox{\boldmath $\zeta$}_k' &=& \mbox{\boldmath $\sigma$}_{k-1}'^{(1)}+\mbox{\boldmath $e$}_k'H_0^T ,
\end{eqnarray}
we have
\begin{equation}
\mbox{\boldmath $\zeta$}_k+\mbox{\boldmath $\zeta$}_k'=(\mbox{\boldmath $\sigma$}_{k-1}^{(1)}+\mbox{\boldmath $\sigma$}_{k-1}'^{(1)})+(\mbox{\boldmath $e$}_k+\mbox{\boldmath $e$}_k')H_0^T .
\end{equation}
These expressions imply that
$$
\mbox{\boldmath $\sigma$}_{k-1}+\mbox{\boldmath $\sigma$}_{k-1}'\mathop{\longrightarrow}_{\mbox{\boldmath $\zeta$}_k+\mbox{\boldmath $\zeta$}_k'}^{\mbox{\boldmath $e$}_k+\mbox{\boldmath $e$}_k'}\mbox{\boldmath $\sigma$}_k+\mbox{\boldmath $\sigma$}_k'
$$
holds.
\end{IEEEproof}
\begin{lemm}Let $\mbox{\boldmath $\beta$}_0$ and $\mbox{\boldmath $\beta$}_N$ be the initial and final states of the code-trellis, respectively. Denote by {\boldmath $y$} a code path connecting these states. (This is denoted as
$$
\mbox{\boldmath $\beta$}_0\stackrel{\mbox{\boldmath $y$}}{\longrightarrow}\mbox{\boldmath $\beta$}_N .\mbox{)}
$$
Then we have
\begin{equation}
\mbox{\boldmath $\beta$}_0^*\mathop{\longrightarrow}_{\mbox{\boldmath $\zeta$}=\mbox{\boldmath $0$}}^{\mbox{\boldmath $y$}}\mbox{\boldmath $\beta$}_N^* .
\end{equation}
That is, assume that the syndrome former is in the dual state $\mbox{\boldmath $\beta$}_0^*$ of $\mbox{\boldmath $\beta$}_0$. In this case, if {\boldmath $y$} is inputted to the syndrome former, then it moves to the dual state $\mbox{\boldmath $\beta$}_N^*$ of $\mbox{\boldmath $\beta$}_N$ and the syndrome $\mbox{\boldmath $\zeta$}=\mbox{\boldmath $0$}$ is outputted.
\end{lemm}
\begin{IEEEproof}By extending the code-trellis in both directions by $L$ sections, if necessary, we can assume the condition
\begin{equation}
\mbox{\boldmath $\beta$}_0=\mbox{\boldmath $0$}\stackrel{\mbox{\boldmath $y$}'}{\longrightarrow}\mbox{\boldmath $\beta$}_L\stackrel{\mbox{\boldmath $y$}}{\longrightarrow}\mbox{\boldmath $\beta$}_{N+L}\stackrel{\mbox{\boldmath $y$}''}{\longrightarrow}\mbox{\boldmath $\beta$}_{N+2L}=\mbox{\boldmath $0$} ,
\end{equation}
where $\mbox{\boldmath $y$}'$ and $\mbox{\boldmath $y$}''$ are augmented code paths (initial and final states are both {\boldmath $0$}). Hence, we can apply the standard scalar parity-check matrix $H_{scalar}$ (cf. (2)). Then we have
\begin{equation}
\mbox{\boldmath $\beta$}_0^*=\mbox{\boldmath $0$}\mathop{\longrightarrow}_{\mbox{\boldmath $\zeta$}'=\mbox{\boldmath $0$}}^{\mbox{\boldmath $y$}'}\mbox{\boldmath $\beta$}_L^*\mathop{\longrightarrow}_{\mbox{\boldmath $\zeta$}=\mbox{\boldmath $0$}}^{\mbox{\boldmath $y$}}\mbox{\boldmath $\beta$}_{N+L}^*\mathop{\longrightarrow}_{\mbox{\boldmath $\zeta$}''=\mbox{\boldmath $0$}}^{\mbox{\boldmath $y$}''}\mbox{\boldmath $\beta$}_{N+2L}^*=\mbox{\boldmath $0$} .
\end{equation}
That is, the output of the syndrome former is zero for all time. In the above relation, we can note the following subsection:
\begin{equation}
\mbox{\boldmath $\beta$}_L^*\mathop{\longrightarrow}_{\mbox{\boldmath $\zeta$}=\mbox{\boldmath $0$}}^{\mbox{\boldmath $y$}}\mbox{\boldmath $\beta$}_{N+L}^* .
\end{equation}
\end{IEEEproof}
\par
Let $\mbox{\boldmath $z$}=\{\mbox{\boldmath $z$}_k\}_{k=1}^N$ be a received data. Denote by $\mbox{\boldmath $\sigma$}_0$ the initial state of the syndrome former. Let $\mbox{\boldmath $\sigma$}_k$ be the syndrome-former state at time $k$ corresponding to the input {\boldmath $z$}. Note that $\mbox{\boldmath $\sigma$}_k$ is independent of $\mbox{\boldmath $\sigma$}_0$ if $k\geq M$. Also, $\mbox{\boldmath $\zeta$}_k$ is independent of $\mbox{\boldmath $\sigma$}_0$ if $k\geq M+1$. In the following, we assume the condition $N\geq M$.
\newtheorem{pro}{Proposition}
\begin{pro}
Let {\boldmath $y$} be a transmitted code path in a tailbiting code subtrellis with $\mbox{\boldmath $\beta$}_0=\mbox{\boldmath $\beta$}_N=\mbox{\boldmath $\beta$}$. Also, let $\mbox{\boldmath $z$}=\mbox{\boldmath $y$}+\mbox{\boldmath $e$}$ be the received data, where {\boldmath $e$} is an error. Denote by $\mbox{\boldmath $\sigma$}_{fin}(=\mbox{\boldmath $\sigma$}_N)$ the final syndrome-former state corresponding to the input {\boldmath $z$}. Here, assume that $\mbox{\boldmath $\sigma$}_0$ is set to $\mbox{\boldmath $\sigma$}_{fin}$ and {\boldmath $z$} is inputted to the syndrome former. Let $\mbox{\boldmath $\zeta$}$ be the outputted syndrome. (Note that the final syndrome-former state is $\mbox{\boldmath $\sigma$}_{fin}$.) Then we have
\begin{equation}
\mbox{\boldmath $\sigma$}_{fin}+\mbox{\boldmath $\beta$}^*\mathop{\longrightarrow}_{\mbox{\boldmath $\zeta$}}^{\mbox{\boldmath $e$}}\mbox{\boldmath $\sigma$}_{fin}+\mbox{\boldmath $\beta$}^* .
\end{equation}
\end{pro}
\begin{IEEEproof}From the assumption, we have
\begin{equation}
\mbox{\boldmath $\sigma$}_{fin}\mathop{\longrightarrow}_{\mbox{\boldmath $\zeta$}}^{\mbox{\boldmath $z$}=\mbox{\boldmath $y$}+\mbox{\boldmath $e$}}\mbox{\boldmath $\sigma$}_{fin} .
\end{equation}
Also, from Lemma 2,
\begin{equation}
\mbox{\boldmath $\beta$}^*\mathop{\longrightarrow}_{\mbox{\boldmath $\zeta$}=\mbox{\boldmath $0$}}^{\mbox{\boldmath $y$}}\mbox{\boldmath $\beta$}^*
\end{equation}
is obtained. Hence, by applying Lemma 1, we have
\begin{equation}
\mbox{\boldmath $\sigma$}_{fin}+\mbox{\boldmath $\beta$}^*\mathop{\longrightarrow}_{\mbox{\boldmath $\zeta$}+\mbox{\boldmath $0$}=\mbox{\boldmath $\zeta$}}^{\mbox{\boldmath $z$}+\mbox{\boldmath $y$}=\mbox{\boldmath $e$}}\mbox{\boldmath $\sigma$}_{fin}+\mbox{\boldmath $\beta$}^* .
\end{equation}
\end{IEEEproof}

\section{Error-Trellises for Tailbiting Convolutional Codes}
\subsection{Error-Trellis Construction}
Suppose that the tailbiting code-trellis based on $G(D)$ is defined in $[0, N]$, where $N\geq M$. In this case, the corresponding tailbiting error-trellis based on $H^T(D)$ is constructed as follows.
\par
{\it Step 1:} Let $\mbox{\boldmath $z$}=\{\mbox{\boldmath $z$}_k\}_{k=1}^N$ be a received data. Denote by $\mbox{\boldmath $\sigma$}_0$ the initial state of the syndrome former $H^T(D)$. Let $\mbox{\boldmath $\sigma$}_{fin}(=\mbox{\boldmath $\sigma$}_N)$ be the final syndrome-former state corresponding to the input {\boldmath $z$}. Note that $\mbox{\boldmath $\sigma$}_{fin}$ is independent of $\mbox{\boldmath $\sigma$}_0$ and is uniquely determined only by {\boldmath $z$}.
\par
{\it Step 2:} Set $\mbox{\boldmath $\sigma$}_0$ to $\mbox{\boldmath $\sigma$}_{fin}$ and input {\boldmath $z$} to the syndrome former. Here, assume that the syndrome sequence $\mbox{\boldmath $\zeta$}=\{\mbox{\boldmath $\zeta$}_k\}_{k=1}^N$ is obtained. ({\it Remark:} $\mbox{\boldmath $\zeta$}_k~(k\geq M+1)$ has been obtained in Step 1.)
\par
{\it Step 3:} Concatenate the error-trellis modules corresponding to the syndromes $\mbox{\boldmath $\zeta$}_k$. Then we have the tailbiting error-trellis.
\par
{\it Example 2:} Again, consider the parity-check matrix $H_1(D)$. Let
\begin{equation}
\mbox{\boldmath $z$}=\mbox{\boldmath $z$}_1~\mbox{\boldmath $z$}_2~\mbox{\boldmath $z$}_3~\mbox{\boldmath $z$}_4~\mbox{\boldmath $z$}_5=111~110~110~111~000
\end{equation}
be the received data. According to Step 1, let us input {\boldmath $z$} to the syndrome former $H_1^T(D)$. Then we have $\mbox{\boldmath $\sigma$}_{fin}=(0,0)$. Next, we set $\mbox{\boldmath $\sigma$}_0$ to $\mbox{\boldmath $\sigma$}_{fin}=(0,0)$ and input {\boldmath $z$} to the syndrome former. In this case, the syndrome sequence
\begin{equation}
\mbox{\boldmath $\zeta$}=\mbox{\boldmath $\zeta$}_1~\mbox{\boldmath $\zeta$}_2~\mbox{\boldmath $\zeta$}_3~\mbox{\boldmath $\zeta$}_4~\mbox{\boldmath $\zeta$}_5=00~00~10~01~11
\end{equation}
is obtained. The tailbiting error-trellis is constructed by concatenating the error-trellis modules corresponding to $\mbox{\boldmath $\zeta$}_k$. The obtained tailbiting error-trellis is shown in Fig.2.

%\begin{figure}[htb]
%\begin{center}
%\includegraphics[width=6.0cm,clip]{tb-2.eps}
%\end{center}
%\caption{Error-trellis modules associated with $H_1^T(D)$.}
%\label{Fig.2}
%\end{figure}
\begin{figure}[tb]
\begin{center}
\includegraphics[width=8.0cm,clip]{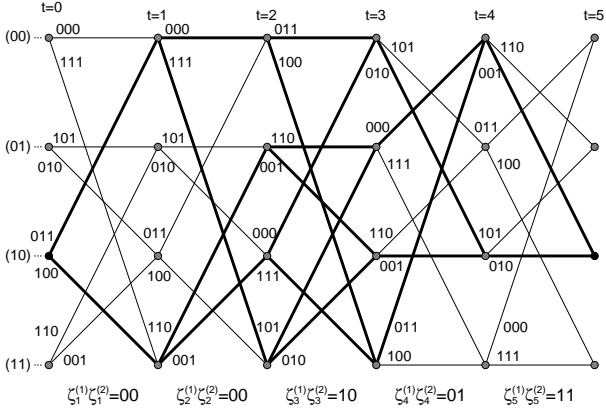}
\end{center}
\caption{Tailbiting error-trellis based on $H_1^T(D)$.}
\label{Fig.2}
\end{figure}

\subsection{Correspondence Between Code Subtrellises and Error Subtrellises}
With respect to the correspondence between tailbiting code subtrellises and tailbiting error subtrellises, we have the following.
\begin{pro}
Let $\mbox{\boldmath $\beta$}_0(=\mbox{\boldmath $\beta$}_N)=\mbox{\boldmath $\beta$}$ be the initial (final) state of a tailbiting code subtrellis. Then the initial (final) state of the corresponding tailbiting error subtrellis is given by $\mbox{\boldmath $\sigma$}_{fin}+\mbox{\boldmath $\beta$}^*$.
\end{pro}
\begin{IEEEproof}Direct consequence of Proposition 1.
\end{IEEEproof}
\par
{\it Example 2 (Continued):} Consider the tailbiting error-trellis in Fig.2. In this example, we have $\mbox{\boldmath $\sigma$}_{fin}=(0,0)$. The corresponding tailbiting code-trellis based on $G_1(D)$ is shown in Fig.1. In Fig.1, take notice of the code subtrellis with initial (final) state $\mbox{\boldmath $\beta$}=(1,0)$ (bold lines). The dual state of $\mbox{\boldmath $\beta$}=(1,0)$ is calculated as $\mbox{\boldmath $\beta$}^*=(u_{-1}+u_0, u_0)=(1+0,0)=(1,0)$. Hence, the initial (final) state of the corresponding error subtrellis is given by $\mbox{\boldmath $\sigma$}_{fin}+\mbox{\boldmath $\beta$}^*=(0,0)+(1, 0)=(1, 0)$ (bold lines in Fig.2).

\subsection{Backward Error-Trellis Construction}
Let $\tilde G(D)$ and $\tilde H(D)$ be the reciprocal encoder and the reciprocal dual encoder [6] associated with $G(D)$, respectively. Then the tailbiting backward error-trellis corresponding to the original tailbiting error-trellis is constructed as follows.
\par
{\it Step 1:} Let $\mbox{\boldmath $\tilde z$}=\{\mbox{\boldmath $\tilde z$}_k\}_{k=1}^N=\{\mbox{\boldmath $z$}_{N-k+1}\}_{k=1}^N$ be the time-reversed received data. Denote by $\mbox{\boldmath $\tilde \sigma$}_0$ the initial state of the syndrome former $\tilde H^T(D)$. Let $\mbox{\boldmath $\tilde \sigma$}_{fin}(=\mbox{\boldmath $\tilde \sigma$}_N)$ be the final syndrome-former state corresponding to the input {\boldmath $\tilde z$}. Note that $\mbox{\boldmath $\tilde \sigma$}_{fin}$ is independent of $\mbox{\boldmath $\tilde \sigma$}_0$ and is uniquely determined only by {\boldmath $\tilde z$}.
\par
{\it Step 2:} Set $\mbox{\boldmath $\tilde \sigma$}_0$ to $\mbox{\boldmath $\tilde \sigma$}_{fin}$ and input {\boldmath $\tilde z$} to the syndrome former. Here, assume that the syndrome sequence $\mbox{\boldmath $\eta$}=\{\mbox{\boldmath $\eta$}_k\}_{k=1}^N$ is obtained.
\par
{\it Remark:} It is shown that $\mbox{\boldmath $\zeta$}=\{\mbox{\boldmath $\zeta$}_k\}_{k=1}^N$ and $\mbox{\boldmath $\eta$}=\{\mbox{\boldmath $\eta$}_k\}_{k=1}^N$ have the following correspondence:
\begin{eqnarray}
\mbox{\boldmath $\eta$} &=& \mbox{\boldmath $\eta$}_1~\mbox{\boldmath $\eta$}_2~\cdots ~\mbox{\boldmath $\eta$}_M~\mbox{\boldmath $\eta$}_{M+1}~\cdots ~\mbox{\boldmath $\eta$}_N \nonumber \\
&=& \mbox{\boldmath $\zeta$}_M~\mbox{\boldmath $\zeta$}_{M-1}~\cdots ~\mbox{\boldmath $\zeta$}_1~\mbox{\boldmath $\zeta$}_N~\cdots ~\mbox{\boldmath $\zeta$}_{M+1} .
\end{eqnarray}
\par
{\it Step 3:} Concatenate the error-trellis modules corresponding to the syndromes $\mbox{\boldmath $\eta$}_k$. Then we have the tailbiting backward error-trellis.
\par
{\it Example 3:} Take notice of Example 2. The reciprocal dual encoder $\tilde H_1(D)$ associated with $G_1(D)$ is given by
\begin{equation}
\tilde H_1(D)=\left(
\begin{array}{ccc}
1+D& 1 & 1+D \\
1& D & D 
\end{array}
\right) .
\end{equation}
Let
\begin{equation}
\mbox{\boldmath $\tilde z$}=\mbox{\boldmath $\tilde z$}_1~\mbox{\boldmath $\tilde z$}_2~\mbox{\boldmath $\tilde z$}_3~\mbox{\boldmath $\tilde z$}_4~\mbox{\boldmath $\tilde z$}_5=000~111~110~110~111
\end{equation}
be the time-reversed received data. According to Step 1, let us input {\boldmath $\tilde z$} to the syndrome former $\tilde H_1^T(D)$. Then we have $\mbox{\boldmath $\tilde \sigma$}_{fin}=(0,0)$. Next, we set $\mbox{\boldmath $\tilde \sigma$}_0$ to $\mbox{\boldmath $\tilde \sigma$}_{fin}=(0,0)$ and input {\boldmath $\tilde z$} to the syndrome former. In this case, the syndrome sequence
\begin{equation}
\mbox{\boldmath $\eta$}=\mbox{\boldmath $\eta$}_1~\mbox{\boldmath $\eta$}_2~\mbox{\boldmath $\eta$}_3~\mbox{\boldmath $\eta$}_4~\mbox{\boldmath $\eta$}_5=00~11~01~10~00
\end{equation}
is obtained. Since $M=1$, we see that the correspondence
\begin{eqnarray}
\mbox{\boldmath $\eta$} &=& \mbox{\boldmath $\eta$}_1~\mbox{\boldmath $\eta$}_2~\mbox{\boldmath $\eta$}_3~\mbox{\boldmath $\eta$}_4~\mbox{\boldmath $\eta$}_5 \nonumber \\
&=& \mbox{\boldmath $\zeta$}_1~\mbox{\boldmath $\zeta$}_5~\mbox{\boldmath $\zeta$}_4~\mbox{\boldmath $\zeta$}_3~\mbox{\boldmath $\zeta$}_2
\end{eqnarray}
holds. The tailbiting backward error-trellis is constructed by concatenating the error-trellis modules corresponding to $\mbox{\boldmath $\eta$}_k$. The obtained tailbiting backward error-trellis is shown in Fig.3.
%\begin{figure}[htb]
%\begin{center}
%\includegraphics[width=6.0cm,clip]{tb-5.eps}
%\end{center}
%\caption{Error-trellis modules associated with $\tilde H_1^T(D)$.}
%\label{Fig.4}
%\end{figure}
\begin{figure}[tb]
\begin{center}
\includegraphics[width=8.0cm,clip]{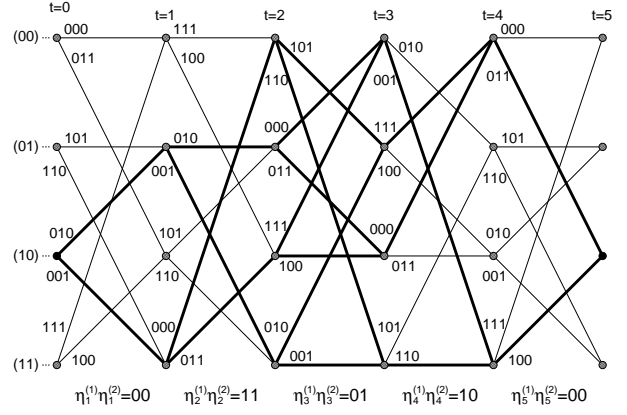}
\end{center}
\caption{Tailbiting backward error-trellis based on $\tilde H_1^T(D)$.}
\label{Fig.3}
\end{figure}
\par
Next, consider the correspondence between forward error subtrellises and backward error subtrellises. First, note the following.
\begin{pro}
Let $\mbox{\boldmath $\tilde \beta$}_0(=\mbox{\boldmath $\tilde \beta$}_N)=\mbox{\boldmath $\tilde \beta$}$ be the initial (final) state of a tailbiting backward code subtrellis. Then the initial (final) state of the corresponding backward error subtrellis is given by $\mbox{\boldmath $\tilde \sigma$}_{fin}+\mbox{\boldmath $\tilde \beta$}^*$.
\end{pro}
\begin{IEEEproof}Direct consequence of Proposition 1.
\end{IEEEproof}
\par
Let $\mbox{\boldmath $\tilde \beta$}$ be the backward state corresponding to $\mbox{\boldmath $\beta$}$. Then the forward code subtrellis with $\mbox{\boldmath $\beta$}_0(=\mbox{\boldmath $\beta$}_N)=\mbox{\boldmath $\beta$}$ and the backward code subtrellis with $\mbox{\boldmath $\tilde \beta$}_0(=\mbox{\boldmath $\tilde \beta$}_N)=\mbox{\boldmath $\tilde \beta$}$ correspond to each other. Hence, using Propositions 2 and 3, we have the following.
\begin{pro}
Let $\mbox{\boldmath $\sigma$}_{fin}+\mbox{\boldmath $\beta$}^*$ be the initial (final) state of a tailbiting forward error subtrellis. Then the initial (final) state of the corresponding backward error subtrellis is given by $\mbox{\boldmath $\tilde \sigma$}_{fin}+\mbox{\boldmath $\tilde \beta$}^*$, where $\mbox{\boldmath $\tilde \beta$}$ is the backward state of $\mbox{\boldmath $\beta$}$.
\end{pro}
\par
{\it Example 3 (Continued):} Consider the reciprocal encoder
\begin{equation}
\tilde G_1(D)=(D^2, 1+D^2, 1+D+D^2)
\end{equation}
and the reciprocal dual encoder $\tilde H_1(D)$ associated with $G_1(D)$. $\tilde H_1(D)$ is expressed as
\begin{eqnarray}
\tilde H_1(D) &=& \left(
\begin{array}{ccc}
1& 1 & 1 \\
1& 0 & 0 
\end{array}
\right)+\left(
\begin{array}{ccc}
1& 0 & 1 \\
0& 1 & 1 
\end{array}
\right)D \nonumber \\
&\stackrel{\triangle}{=}& \tilde H_0+\tilde H_1D .
\end{eqnarray}
Hence, the dual state corresponding to $\mbox{\boldmath $\tilde \beta$}_k=(u_{k-1}, u_k)$ is calculated as
\begin{eqnarray}
\mbox{\boldmath $\tilde \beta$}_k^* &=& \mbox{\boldmath $y$}_k\tilde H_1^T \nonumber \\
&=& (y_k^{(1)}, y_k^{(2)}, y_k^{(3)}) \left(
\begin{array}{cc}
1 & 0 \\
0 & 1 \\
1 & 1 
\end{array}
\right) \nonumber \\
&=& (y_k^{(1)}+y_k^{(3)}, y_k^{(2)}+y_k^{(3)}) \nonumber \\
&=& (u_{k-1}+u_k, u_{k-1}) .
\end{eqnarray}
Here, take notice of the error subtrellis with initial (final) state $(1, 0)$ in Fig.2. (Note that $\mbox{\boldmath $\sigma$}_{fin}+\mbox{\boldmath $\beta$}^*=(0,0)+(1, 0)=(1, 0)$.) This error subtrellis corresponds to the code subtrellis with initial (final) state $\mbox{\boldmath $\beta$}=(1,0)$ in Fig.1. On the other hand, the backward state of $\mbox{\boldmath $\beta$}=(1,0)$ is $\mbox{\boldmath $\tilde \beta$}=(0,1)$ and its dual state becomes $\mbox{\boldmath $\tilde \beta$}^*=(u_{-1}+u_0, u_{-1})=(0+1,0)=(1,0)$. Hence, from Proposition 4, the initial (final) state of the corresponding backward error subtrellis is given by $\mbox{\boldmath $\tilde \sigma$}_{fin}+\mbox{\boldmath $\tilde \beta$}^*=(0,0)+(1,0)=(1,0)$ (bold lines in Fig.3).

\section{$H_{scalar}$ for Tailbiting Convolutional Codes}
Consider the tailbiting convolutional code $C_{tb}$ with $N$ trellis sections specified by a parity-check matrix $H(D)$. $C_{tb}$ can be regarded as an $(Nn, Nk)$ block code [4]. In this case, we have the following.
\begin{pro} Assume that $H(D)$ has the form (1). Then the scalar parity-check matrix $H_{scalar}$ for $C_{tb}$ is given by
\begin{equation}
\arraycolsep=1pt
\left(
\begin{array}{cccccccc}
H_0 &  &  &  & H_M & \scriptstyle{\ldots} & H_2 & H_1 \\
H_1 & H_0 &  &  &  & \scriptstyle{\ldots} & \scriptstyle{\ldots} & H_2 \\
\scriptstyle{\ldots} & H_1 & \scriptstyle{\ldots} &  &  &  & H_M & \scriptstyle{\ldots} \\
H_{M-1} & \scriptstyle{\ldots} & \scriptstyle{\ldots} & H_0 &  &  &  & H_M \\
H_M & H_{M-1} & \scriptstyle{\ldots} & H_1 & H_0 &  &  &  \\
 & H_M & \scriptstyle{\ldots} & \scriptstyle{\ldots} & H_1 & \scriptstyle{\ldots} &  &  \\
 &  & \scriptstyle{\ldots} & H_{M-1} & \scriptstyle{\ldots} & \scriptstyle{\ldots} & H_0 &  \\
 &  &  & H_M & H_{M-1} & \scriptstyle{\ldots} & H_1 & H_0
\end{array}
\right)
\end{equation}
with size $Nr \times Nn$.
\end{pro}
\begin{IEEEproof}Consider the tailbiting error-trellis of $C_{tb}$. It is characterized by the condition $\mbox{\boldmath $\sigma$}_0=\mbox{\boldmath $\sigma$}_N$. Accordingly, we have the following equalities.
\begin{eqnarray}
&& \mbox{\boldmath $e$}_{-M+1}H_M^T+ \cdots +\mbox{\boldmath $e$}_{-1}H_2^T+\mbox{\boldmath $e$}_0H_1^T \nonumber \\
&=& \mbox{\boldmath $e$}_{N-M+1}H_M^T+ \cdots +\mbox{\boldmath $e$}_{N-1}H_2^T+\mbox{\boldmath $e$}_NH_1^T \\
&& \mbox{\boldmath $e$}_{-M+2}H_M^T+ \cdots +\mbox{\boldmath $e$}_{-1}H_3^T+\mbox{\boldmath $e$}_0H_2^T \nonumber \\
&=& \mbox{\boldmath $e$}_{N-M+2}H_M^T+ \cdots +\mbox{\boldmath $e$}_{N-1}H_3^T+\mbox{\boldmath $e$}_NH_2^T \\
&& \cdots \nonumber \\
&& \mbox{\boldmath $e$}_0H_M^T=\mbox{\boldmath $e$}_NH_M^T .
\end{eqnarray}
Hence, the syndrome $\mbox{\boldmath $\zeta$}_1$ is expressed as
\begin{eqnarray}
\mbox{\boldmath $\zeta$}_1 &=& (\mbox{\boldmath $e$}_{-M+1}H_M^T+ \cdots +\mbox{\boldmath $e$}_{-1}H_2^T+\mbox{\boldmath $e$}_0H_1^T)+\mbox{\boldmath $e$}_1H_0^T \nonumber \\
&=& (\mbox{\boldmath $e$}_{N-M+1}H_M^T+ \cdots +\mbox{\boldmath $e$}_{N-1}H_2^T+\mbox{\boldmath $e$}_NH_1^T)+\mbox{\boldmath $e$}_1H_0^T \nonumber \\
&=& (\mbox{\boldmath $e$}_1, \mbox{\boldmath $e$}_2, \cdots, \mbox{\boldmath $e$}_{N-M+1},\cdots, \mbox{\boldmath $e$}_N) \nonumber \\
&& \times (H_0, 0, \cdots, 0, H_M, \cdots, H_1)^T .
\end{eqnarray}
Similarly, we have
\begin{eqnarray}
\mbox{\boldmath $\zeta$}_2 &=& (\mbox{\boldmath $e$}_{-M+2}H_M^T+ \cdots +\mbox{\boldmath $e$}_0H_2^T)+\mbox{\boldmath $e$}_1H_1^T+\mbox{\boldmath $e$}_2H_0^T \nonumber \\
&=& (\mbox{\boldmath $e$}_{N-M+2}H_M^T+ \cdots +\mbox{\boldmath $e$}_NH_2^T)+\mbox{\boldmath $e$}_1H_1^T+\mbox{\boldmath $e$}_2H_0^T \nonumber \\
&=& (\mbox{\boldmath $e$}_1, \mbox{\boldmath $e$}_2, \cdots, \mbox{\boldmath $e$}_{N-M+2},\cdots, \mbox{\boldmath $e$}_N) \nonumber \\
&& \times (H_1, H_0, 0, \cdots, 0, H_M, \cdots, H_2)^T .
\end{eqnarray}
The same argument can be applied to $\mbox{\boldmath $\zeta$}_k~(3\leq k \leq N)$. Then we see that $H_{scalar}^T$ is written as
\begin{equation}
\arraycolsep=1pt
\left(
\begin{array}{cccccccc}
H_0^T & H_1^T & \scriptstyle{\ldots} & H_{M-1}^T & H_M^T &  &  &  \\
 & H_0^T & \scriptstyle{\ldots} & \scriptstyle{\ldots} & H_{M-1}^T & H_M^T &  &  \\
 &  & \scriptstyle{\ldots} & H_1^T & \scriptstyle{\ldots} & H_{M-1}^T & \scriptstyle{\ldots} &  \\
 &  &  & H_0^T & H_1^T & \scriptstyle{\ldots} & \scriptstyle{\ldots} & H_M^T \\
H_M^T &  &  &  & H_0^T & H_1^T & \scriptstyle{\ldots} & H_{M-1}^T \\
H_{M-1}^T & H_M^T &  &  &  & H_0^T & \scriptstyle{\ldots} & \scriptstyle{\ldots} \\
\scriptstyle{\ldots} & \scriptstyle{\ldots} & \scriptstyle{\ldots} &  &  &  & \scriptstyle{\ldots} & H_1^T \\
H_1^T & H_2^T & \scriptstyle{\ldots} & H_M^T &  &  &  & H_0^T
\end{array}
\right) .
\end{equation}
By transposing this matrix, $H_{scalar}$ is obtained.
\end{IEEEproof}

\section{Conclusion}
In this paper, we have presented an error-trellis construction for tailbiting convolutional codes. A tailbiting error-trellis is characterized by the condition that the syndrome former starts and ends in the same state. We have clarified the correspondence between code subtrellises in the tailbiting code-trellis and error subtrellises in the tailbiting error-trellis. Also, we have presented a construction of tailbiting backward error-trellises. Moreover, we have obtained the general structure of the scalar parity-check matrix for a tailbiting convolutional code. We see that the obtained results correspond to those for tailbiting code-trellises in the natural manner.


%\section*{Acknowledgment}




%\bibliographystyle{IEEEtran}

\end{document}